\documentclass[conference]{IEEEtran}
\IEEEoverridecommandlockouts

\usepackage{cite}
\usepackage{amsmath,amssymb,amsfonts}
\usepackage{algorithmic}
\usepackage{graphicx}
\usepackage{textcomp}
\usepackage{url}
\usepackage{booktabs}
\usepackage{xcolor}
\def\BibTeX{{\rm B\kern-.05em{\sc i\kern-.025em b}\kern-.08em
    T\kern-.1667em\lower.7ex\hbox{E}\kern-.125emX}}
\begin{document}

\title{Incorporating Spatial Cues in Modular Speaker Diarization for Multi-channel Multi-party Meetings
}

\author{
    \IEEEauthorblockN{
        Ruoyu Wang\IEEEauthorrefmark{1},
        Shutong Niu\IEEEauthorrefmark{1},
        Gaobin Yang\IEEEauthorrefmark{1},
        Jun Du\IEEEauthorrefmark{1}\IEEEauthorrefmark{2}\thanks{\IEEEauthorrefmark{2} Corresponding author},
        Shuangqing Qian\IEEEauthorrefmark{3},
        Tian Gao\IEEEauthorrefmark{3}
        Jia Pan\IEEEauthorrefmark{3}
    }
    \IEEEauthorblockA{\IEEEauthorrefmark{1}NERC-SLIP, University of Science and Technology of China (USTC), Hefei, China}
    \IEEEauthorblockA{\IEEEauthorrefmark{3}iFlytek Research, Hefei, China}
    \{wangruoyu, niust, gaobinyang\}@mail.ustc.edu.cn, jundu@ustc.edu.cn
}


\maketitle

\begin{abstract}
Although fully end-to-end speaker diarization systems have made significant progress in recent years, modular systems often achieve superior results in real-world scenarios due to their greater adaptability and robustness. Historically, modular speaker diarization methods have seldom discussed how to leverage spatial cues from multi-channel speech. This paper proposes a three-stage modular system to enhance single-channel neural speaker diarization systems and recognition performance by utilizing spatial cues from multi-channel speech to provide more accurate initialization for each stage of neural speaker diarization (NSD) decoding: (1) Overlap detection and continuous speech separation (CSS) on multi-channel speech are used to obtain cleaner single speaker speech segments for clustering, followed by the first NSD decoding pass. (2) The results from the first pass initialize a complex Angular Central Gaussian Mixture Model (cACGMM) to estimate speaker-wise masks on multi-channel speech, and through Overlap-add and Mask-to-VAD, achieve initialization with lower speaker error (SpkErr), followed by the second NSD decoding pass. (3) The second decoding results are used for guided source separation (GSS), recognizing and filtering short segments containing less one word to obtain cleaner speech segments, followed by re-clustering and the final NSD decoding pass. We presented the progressively explored evaluation results from the CHiME-8 NOTSOFAR-1 (Natural Office Talkers in Settings Of Far-field Audio Recordings) challenge, demonstrating the effectiveness of our system and its contribution to improving recognition performance. Our final system achieved the first place in the challenge.
\end{abstract}

\begin{IEEEkeywords}
CHiME challenge, speaker diarization, speech separation, multi-channel multi-party meetings
\end{IEEEkeywords}

\section{Introduction}
The task of speaker diarization in multi-party meetings plays a crucial role in speech processing, tackling the challenge of determining ``who spoke when" in conversations involving multiple participants~\cite{park2022review}.
Accurate speaker diarization is vital for enhancing the performance of automatic speech recognition (ASR) systems, facilitating precise meeting transcriptions, and enabling a deeper analysis of conversational dynamics~\cite{anguera2012speaker}.
The importance of this task is further emphasized by several high-profile competitions, such as DIHARD-III~\cite{ryant2020third}, CHiME-7~\cite{cornell2023chime}, and M2MeT~\cite{yu2022m2met}, which have propelled significant advancements and innovations in the field. 

Neural network-based speaker diarization methods are generally classified into modular systems and end-to-end systems. Modular systems typically start with Voice Activity Detection (VAD) for initial segmentation, followed by clustering to estimate the number of speakers and assign speaker labels~\cite{shum2013unsupervised}. These results are then refined using deep learning models, such as Target-Speaker Voice Activity Detection (TS-VAD)~\cite{medennikov2020stc,he2021target}. 
In contrast, end-to-end neural speaker diarization (EEND)~\cite{fujita2019end,horiguchi2020end}, aim to jointly estimate speaker labels and speech activity in a unified process. 
However, end-to-end methods encounter challenges in real-world scenarios, such as difficulties in controlling the number of speakers and performance degradation due to speaker permutation error. Conversely, modular systems provide notable advantages, including greater adaptability and robustness, especially in handling varying numbers of speakers and complex acoustic environments. Importantly, modular systems have consistently demonstrated superior performance across various real-world datasets, highlighting their effectiveness and reliability in practical applications~\cite{ryant2020third,cornell2023chime,yu2022m2met}.

In modular speaker diarization systems, leveraging spatial information from multi-channel data has become a significant research focus. Various teams have explored using spatial information from multi-channel data to enhance diarization results and speech recognition performance in competitions such as CHiME-6~\cite{watanabe2020chime}, CHiME-7~\cite{cornell2023chime}, M2MeT~\cite{yu2022m2met}. 
For example, in CHiME-6, the USTC team used Neural Beamforming for SS (NBSS) to enhance traditional GSS, resulting in cleaner segments and improved VAD performance~\cite{gaoustc}. 
In CHiME-7, the USTC team applied an iterative cACGMM-based diarization correction method, achieving progressively better results through a four-stage process~\cite{wang2023ustc}. 
Moreover, the top-performing system in M2MeT utilized a multi-channel TS-VAD model~\cite{wang2022cross}, while the CUHK-TENCENT team explored the use of DOA methods to estimate speaker locations and integrate this spatial information into Neural Speaker Diarization (NSD) models~\cite{zheng2022cuhk}.

However, these methods have limitations. Typically, the number of participants in the datasets is either fixed or varies only slightly, which restricts the exploration of spatial cues in more dynamic settings.
Moreover, distinguishing the specific contributions of spatial cues can be challenging with multi-channel deep learning models, as these models often integrate features in complex ways. 
This highlights the need for more extensive research into effectively utilizing multi-channel information, especially in scenarios with a larger number of participants and more variable conditions.

The latest CHiME-8~\cite{cornell2024chime} NOTSOFAR-1 (Natural Office Talkers in Settings Of Far-field Audio Recordings) Challenge~\cite{vinnikov2024notsofar} focuses on distant speaker diarization and automatic speech recognition using a single recording device, emulating realistic conversational environments. 
This challenge introduces a benchmarking dataset comprising approximately 280 distinct meetings recorded in various conference rooms, along with a corresponding simulated training dataset designed to address previous train-test gaps. 
The benchmarking dataset captures a wide range of real-world conditions, including speech near whiteboards, speakers at varying distances and volumes, transient and persistent noise, and complex meeting dynamics such as overlapping speech and rapid speaker changes. 
With recordings from 30 rooms of different sizes, layouts, and acoustic characteristics, and involving 4-8 participants per meeting, this dataset presents new challenges for the development and evaluation of speaker diarization systems.

To address these challenges, our work introduces a three-stage modular speaker diarization system that progressively enhances single-channel neural speaker diarization (NSD) decoding by leveraging spatial cues from multi-channel data:
in Stage 1, we first apply a multi-channel overlap model to detect overlapping and non-overlapping speech segments. For non-overlapping segments, we use beamforming techniques; for overlapping segments, we employ multi-channel Continuous Speech Separation (CSS) methods. This approach yields cleaner single-speaker audio segments for clustering, thereby improving the initial NSD decoding results.
In Stage 2, we further refine the diarization results using the cACGMM rectification method proposed in CHiME-7~\cite{wang2023ustc,ma2024spatial}. 
The results from the first decoding pass are used to initialize the Complex Angular Center Gaussian Mixture Model (cACGMM)~\cite{ito2016complex} for speaker mask estimation on multi-channel data. This is followed by overlap-add abd mask-to-VAD conversion, which provides lower speaker error (SpkErr) initialization for a second pass of NSD decoding.
In Stage 3, we apply Guided Source Separation (GSS)~\cite{raj2022gpu} using the results from Stage 2 to identify and filter out short segments containing one or fewer words, thereby obtaining clearer speaker segments for more accurate re-clustering and a third pass of NSD decoding.
We report the speaker diarization results and corresponding recognition performance of the three-stage system across various scenarios in the NOTSOFAR-1 dataset~\cite{vinnikov2024notsofar}, demonstrating the effectiveness of our approach and its contribution to improving recognition outcomes.

\section{System Description}
\begin{figure}[t]
\centering
\includegraphics[width=1.0\linewidth]{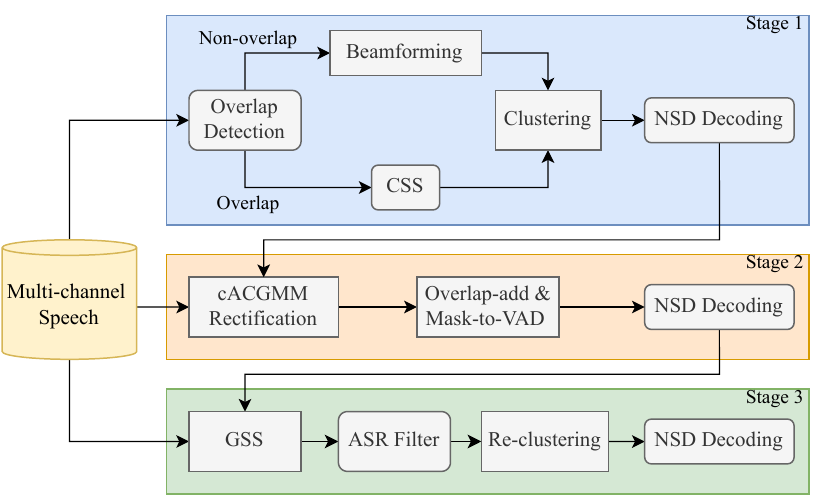}
\caption{The pipeline for the proposed three-stage modular speaker diarization.}
\label{M1}
\end{figure}

To fully leverage the spatial information provided by multi-channel data, we propose a three-stage modular speaker diarization system. As shown in Figure~\ref{M1}, this system is designed as a tandem approach, incrementally enhancing performance by integrating multi-channel speech separation, single-channel neural speaker diarization (NSD), and traditional multi-channel blind source separation techniques. Each stage refines the NSD decoding by optimizing the speaker activity information provided by the previous stage, leading to more precise initialization for NSD decoding. For the original multi-channel data, we default to applying the Weighted Prediction Error (WPE)~\cite{drude2018nara} algorithm for speech dereverberation. 

\subsection{Stage 1: CSS Segments Clustering Initialization}

We first perform overlap segment detection. The overlap segment detection system uses the same Conformer architecture as the official CSS baseline multi-channel separation model~\cite{vinnikov2024notsofar}. 
We modified the sliding window length to 800 frames (12.8 seconds) and adjusted the final prediction output to frame-level binary classification using a linear layer.

For detected overlapping segments, we applied the multi-channel continuous speech separation (CSS) method to effectively separate each speaker’s voice. To improve performance, we modified the official baseline CSS architecture by adding a classification network for overlap detection and performed joint training for both separation and overlap segment detection. The official model was used for parameter initialization, followed by joint training to enhance the separation model's ability to distinguish between overlapping and non-overlapping segments. We utilized only the prediction results from the separation module, with a sliding window length of 3 seconds. The training and inference procedures remained consistent with the official baseline, using the same training data and relying solely on the official simulated dataset. For non-overlapping segments, we employed an MVDR~\cite{souden2010study} beamformer to enhance the multi-channel speech.

Subsequently, we applied a clustering-based speaker diarization (CSD) method to the preprocessed speech to generate preliminary speaker distribution priors. 
The spectral clustering algorithm and the VAD model from the 3D-speaker toolkit~\cite{chen20243d} were employed, and speaker embeddings were extracted using a ResNet-221~\cite{he2016deep} model trained on the VoxCeleb~\cite{nagrani2017voxceleb} and LibriSpeech~\cite{panayotov2015librispeech} datasets. The clustering results were used as initial priors and fed into the NSD system to obtain more precise speaker boundary information.

\subsection{Stage 2: cACGMM Rectification Initialization}

In the second stage, we apply the complex Angular Central Gaussian Mixture Model (cACGMM) rectification to the NSD decoding results from Stage 1 to generate new initialization. First, the multi-channel speech $\boldsymbol{Y}$ and NSD decoding results $\boldsymbol{D}$ are segmented into long overlapping blocks. Each block is then processed by cACGMM to estimate the presence probabilities of each speaker and noise, represented as speaker masks. By thresholding these masks and using the overlap-add method, we concatenate the speaker masks from all blocks into long sequential masks, generating voice activity detection (VAD) for each speaker. This method allows us to refine the spatial information across all channels, leading to improved single-channel NSD results.

Sequential inputs $\boldsymbol{Y}$ and $\boldsymbol{D}$ are divided into blocks of length $B$ (pre-defined) with a hop size of $W = B/2$, resulting in 50\% overlap. The final block length is $L - \left\lfloor L/W \right\rfloor \times W$, where $\left\lfloor \cdot \right\rfloor$ is the floor function. Each block's inputs are denoted as $\boldsymbol{Y}_{s}$ and $\boldsymbol{D}_{s}$ for $s = 1,...,S$, with $S = \left\lceil L/W \right\rceil$ blocks, where $\left\lceil \cdot \right\rceil$ is the ceiling function. The block size $B$ must be sufficiently large to ensure adequate speech coverage for most of speakers. Segmentation outputs $\boldsymbol{Y}_{s}$ and $\boldsymbol{D}_{s}$ are then passed to the cACGMM with the cACG distribution $\mathcal{A}(\mathbf{z}; \mathbf{B})$~\cite{kent1997data}:
\begin{align}
\label{eq:cacgmm1}
p(\mathbf{z}(l,f)) = \sum_{k=1}^{K+1} \pi(l,f) \mathcal{A}(\mathbf{z}(l,f); \mathbf{B}(l,f)),
\end{align}
where $\pi (l,f)$ is a time-varying priori probability and $\mathbf{z}(l,f) = \mathbf{Y}(l,f) / || \mathbf{Y}(l,f) ||$. 
The total number of classes, $K+1$, includes $K$ speakers plus one for noise. The cACGMM parameters are optimized using the EM algorithm, alternating between E- and M-steps. The update equations are as follows:
\begin{align}
\label{eq:cacgmm2}
\resizebox{0.9\hsize}{!}{$\gamma (l,f,k) = \frac{\pi (l,f)\mathrm{det} (\mathbf{B}(l,f))^{-1} (\mathbf{z}(l,f)^{\mathrm{H} } \mathbf{B}(l,f)^{-1} \mathbf{z}(l,f))^{-N} }{\sum_{k=1}^{K+1}\pi (l,f)\mathrm{det} (\mathbf{B}(l,f))^{-1} (\mathbf{z}(l,f)^{\mathrm{H} } \mathbf{B}(l,f)^{-1} \mathbf{z}(l,f))^{-N}}$}, \\
\label{eq:cacgmm3}
\resizebox{0.9\hsize}{!}{$\mathbf{B}(l,f) = \frac{N}{\sum_{l=1}^{L} \gamma (l,f,k)} \sum_{l=1}^{L} \gamma (l,f,k)\frac{\mathbf{z}(l,f)^{\mathrm{H}}\mathbf{z}(l,f)}{\mathbf{z}(l,f)^{\mathrm{H}}  \mathbf{B}(l,f)^{-1} \mathbf{z}(l,f)}$},
\end{align}
where $\gamma (l,f,k)$ is the posterior probability that the class $k$ is active at the time-frequency $(l,f)$ bin. 

The cACGMM algorithm's EM procedure converges to a local optimum, making it sensitive to initialization. Therefore, selecting appropriate initial values for the posterior $\gamma(l,f,k)$ is crucial.
For block $s$, the posterior is initialled with the diarization results $d_{s}(l,k)$ as
\begin{eqnarray}
\label{eq:initp}
\gamma_{s}^{\mathrm{init} } (l,f,k) = \frac{d_{s}(l,f,k)}{\sum_{k=1}^{K+1} d_{s}(l,f,k)},
\end{eqnarray}
where $d_{s}(l,f,k)$ is a copy of $d_{s}(l,k)$ in the frequency dimension. 
Assuming noise is present at all times, with $d_{s}(t, K+1) = 1$ for $t=1,...,T$, current deep learning-based novelty diarization algorithms have limited research on utilizing spatial information in multi-array scenarios. This is due to the difficulty in constructing a diverse training dataset with various array shapes and distributions. The cACGMM algorithm, being adaptive to test signals, can leverage spatial information at each time step. If a speaker is misidentified or missed in a long-term block, the cACGMM model can correct these errors through iteration, significantly reducing speaker errors, particularly in low SNR conditions or among speakers of the same gender.

After the EM algorithm converges with initialization from diarization results in each block, the speech presence probability at the $(l,f)$ bin is determined using the learned class posterior probabilities, given by:
\begin{eqnarray}
\label{eq:dsre}
\resizebox{0.9\hsize}{!}{
$\gamma_{s}^\mathrm{re}(l,f,k) =\left\{\begin{matrix}
  & \gamma^\mathrm{dia}_{s} (l,f,k), & s=1  \\
  & [\gamma^\mathrm{dia}_{s-1} (l,f,k)+\gamma^\mathrm{dia}_{s} (l,f,k)]/2, & s>1,
\end{matrix}\right.$
}
\end{eqnarray}
where $(s-1)W \le l < sW$. 
All blocks are then concatenated together to form a 3-D tensor $\boldsymbol{\gamma}^\mathrm{re}=[\boldsymbol{\gamma}_{1}^\mathrm{re}, ..., \boldsymbol{\gamma}_{S}^\mathrm{re}] \in \mathbb{R} ^{L\times F\times K}$. The final speaker activity at frame level, denoted as $M_{\mathrm{vad}}^\mathrm{re}(l,k)$, which also provides the diarization information of which speaker is active and when can be obtained as follow:
\begin{eqnarray}
\label{eq:dsre1}
\resizebox{0.9\hsize}{!}{
$M_{\mathrm{vad}}^\mathrm{re}(l,k)=\left\{\begin{matrix}
  & 1, &\text{ if } \beta (l,k) > 0.2, l=l-6,...,l \\
  & 0, & else,
\end{matrix}\right.$
}
\end{eqnarray}
where $\beta (l,k) = (\sum_{f=1}^{F} {\gamma}^\mathrm{re}(l,f,k))/F$ can provide the  probability of speaker presence at each frame $l$. We also can repeat the whole process using the refined speaker activity 
$\boldsymbol{M}_{\mathrm{vad}}^\mathrm{re}$.

\subsection{Stage 3: GSS Segments Re-clustering Initialization}
In Stage 3, we initialize the Guided Source Separation (GSS)~\cite{raj2022gpu} using the NSD decoding results from Stage 2 to obtain separated segments corresponding to the speech activity of each speaker. 
NSD results from Stage 2 often produce more short segments compared to the VAD results from Stage 1. 
Since GSS separation and clustering algorithms struggle with very short segments, we first filter out segments containing less one word using a speech recognition model to avoid interference from incomplete or extremely short segments. 
Subsequently, we perform re-clustering using the spectral clustering algorithm to refine the speaker priors, further reducing speaker errors, and proceed with the final NSD decoding.

\section{Experiment}
\subsection{Experimental Setup}

The neural speaker diarization (NSD) model we used, memory-aware multi-speaker embedding with sequence-to-sequence architecture (NSD-MS2S)~\cite{yang2024neural}, demonstrated excellent performance in the CHiME-7.
Since the Train-1, Train-2, and Dev-1 datasets from the official release come from the same scenario with overlapping speakers, we first train the NSD model \(M_{simu}\) on simulated datasets to avoid overfitting during the method exploration phase. 
This includes 7-channel simulated meetings from the official dataset, as well as clean and noisy meetings simulated from LibriSpeech~\cite{panayotov2015librispeech} (using MUSAN~\cite{snyder2015musan} noise and noise from the official simulated dataset) by jsalt2020-simulate\footnote{\url{https://github.com/jsalt2020-asrdiar/jsalt2020_simulate}}, totaling 5000 hours of 7-channel meetings.

Next, we load the pre-trained parameters of \(M_{simu}\) and fine-tune them using Train-1, Train-2, Dev-1, and far-field multi-channel data simulated with near-field data with near-field data and the noises from official data, resulting in the model \(M_{tune}\). 
This model is tested on Dev-2 and Eval to evaluate the effectiveness of the three-stage approach on specific datasets.

The accuracy of the speaker diarization system is measured by the diarization error rate (DER)~\cite{fiscus2006rich}, calculated as the sum of three types of errors—false alarm (FA), missed detection (MISS), and speaker errors (SpkErr)—divided by the total duration time. ASR results are evaluated using the speaker-attributed time-constrained minimum permutation WER (tcpWER)~\cite{von2023meeteval} metric to assess the impact of speaker diarization errors and word errors.

\subsection{Results and Analysis}

\begin{table}[t]
\centering
\caption{
Evaluation results of the proposed three-stage method on Train-1 and Train-2. The NSD model uses \(M_{simu}\), and the ASR model employs the open-source Whisper-large-v3
}
\setlength{\tabcolsep}{3pt} 
\resizebox{\columnwidth}{!}{
\begin{tabular}{l|cccc|cccc|c}
\toprule
 & \multicolumn{4}{c|}{Initialization} & \multicolumn{4}{|c|}{NSD Decoding} & \multicolumn{1}{c}{ASR} \\
       & FA   & MISS  & SpkErr & DER   & FA   & MISS  & SpkErr & DER   & tcpWER \\
\midrule
Oracle                     & -    & -    & -      & -     & 3.49 & 9.76  & 3.03   & 16.28 & 17.62      \\
Stage 1 (w/o CSS)           & 3.68 & 27.36 & 3.33   & 34.37 & 2.96 & 11.68 & 3.76   & 18.39 & -      \\
Stage 1                    & 3.28 & 21.06 & 2.99   & 27.34 & 4.02 & 9.85  & 3.73   & 17.59 & 19.58  \\
Stage 2                    & 4.76 & 13.23 & 1.89   & 19.88 & 3.22 & 10.53 & 3.72   & 17.47 & 19.04  \\
Stage 3 (w/o filter)        & 3.08 & 11.69 & 2.63   & 17.40 & 3.48 & 10.32 & 3.39   & 17.19 & 18.04  \\
\bottomrule
\end{tabular}}
\label{t1}
\end{table}
As shown in Table~\ref{t1}, in Stage 1, clustering is performed on speech separated using CSS. Compared to directly clustering the speech from channel 0 without processing, performance improves from 34.37 to 27.34, with reductions in FA, MISS, and SpkErr.
In Stage 2, after applying cACGMM rectification, the overall DER improves from Stage 1 initialization, with reductions in MISS and SpkErr, although FA increases. This is due to the need for overlap-add before converting the Stage 2 mask into VAD, which helps prevent missing speaker information at the boundaries.
Since the model is trained on simulated data and the official training set has not been used, we employed the open-source Whisper-large-v3~\cite{radford2023robust} model without dataset adaptation. 
Consequently, in Stage 3, we did not apply word filtering during re-clustering to avoid mistakenly discarding segments due to adaptation issues, which could negatively impact clustering performance.
After reclustering, further improvements are observed compared to Stage 1, with enhancements in FA, MISS, and SpkErr. Throughout each stage, both our NSD decoding and recognition performance progressively improve, approaching the oracle decoding. 

However, after training the NSD model with test-adapted training data, the performance in Stage 2 and 3 changes. 
Table~\ref{t2} shows that the trends in the first two stages are the same as the simulated training set. In Stage 3, skipping word filtering and proceeding with direct reclustering results in a sharp increase in initialization SpkErr, which adversely affects subsequent decoding.
Comparing the oracle decoding results in Tables~\ref{t1} and~\ref{t2}, incorporating real data to train the NSD system significantly reduces MISS, indicating improved segment detection. However, in the NOTSOFAR-1 dataset, many MISS errors are due to short segments that overlap entirely with other speakers. For instance, when Speaker 1 presents ideas, Speaker 2 might interject with phrases like ``OK" or ``Right". 
These fully overlapping short segments, even if detected during NSD decoding, cannot be effectively separated by GSS, leading to degraded re-clustering performance.
After applying word filtering in Stage 3, SpkErr decreases further. Although MISS increases due to the removal of short segments with only one word, NSD decoding does not show additional improvement relative to Stage 1. However, ASR results reflect an improvement in tcpWER.

Table~\ref{t3} shows that the NSD decoding and recognition trends across the three stages are consistent with Table~\ref{t2}.
The best speaker diarization results are achieved with NSD decoding in Stage 2. In Stage 3, due to the word filtering technique, MISS increases during initialization, but SpkErr is reduced compared to Stage 1. The Stage 3 NSD decoding results show lower FA and SpkErr. The best recognition results are achieved in Stage 3, likely due to the style of the meeting, where MISS errors, especially the short segments filtered out by our process, have less impact on recognition performance.
\begin{table}[t]
\centering
\caption{
Evaluation results of the proposed three-stage method on Dev-2 (w/o session 30552). The NSD model uses \(M_{tune}\), and the ASR model is based on Whisper-large-v3, fine-tuned on Train-1 and Train-2~\cite{niu2024ustc}
}
\setlength{\tabcolsep}{3pt} 
\resizebox{\columnwidth}{!}{
\begin{tabular}{l|cccc|cccc|c}
\toprule
& \multicolumn{4}{c|}{Initialization} & \multicolumn{4}{|c|}{NSD Decoding} & ASR \\
       & FA   & MISS  & SpkErr & DER   & FA   & MISS  & SpkErr & DER   & tcpWER \\
\midrule
Oracle                     & -    & -    & -      & -     & 3.50 & 6.34  & 1.25   & 11.10 & 11.37      \\
Stage 1 (w/o CSS)           & 4.72 & 25.36 & 2.56   & 32.65 & 4.82 & 7.46  & 2.96   & 15.24 & -      \\
Stage 1                    & 5.90 & 15.17 & 2.36   & 23.43 & 4.77 & 7.37  & 2.26   & 14.40 & 12.87  \\
Stage 2                    & 7.51 & 11.67 & 1.89   & 21.07 & 4.51 & 7.00  & 2.47   & 13.97 & 14.13  \\
Stage 3 (w/o filter)        & 3.50 & 8.38  & 4.25   & 16.12 & 4.13 & 7.44  & 3.00   & 14.58 & 13.65  \\
Stage 3           & 3.71 & 13.87 & 1.50   & 19.09 & 4.50 & 7.47  & 2.21   & 14.19 & 12.83  \\
\bottomrule
\end{tabular}}
\label{t2}
\end{table}

\begin{table}[t]
\centering
\caption{Evaluation results of the proposed three-stage method on Eval. The NSD model uses \(M_{tune}\), and the ASR model is based on Whisper-large-v3, fine-tuned on Train-1 and Train-2~\cite{niu2024ustc}}
\setlength{\tabcolsep}{3pt} 
\resizebox{\columnwidth}{!}{
\begin{tabular}{l|cccc|cccc|c}
\toprule
& \multicolumn{4}{c|}{Initialization} & \multicolumn{4}{|c|}{NSD Decoding} & ASR \\
       & FA   & MISS  & SpkErr & DER   & FA   & MISS  & SpkErr & DER   & tcpWER \\
\midrule
Stage 1                    & 3.77 & 15.55 & 2.64   & 21.96 & 3.44 & 8.21  & 3.06   & 14.71 & 11.76      \\
Stage 2                    & 2.88 & 12.76 & 0.88   & 16.52 & 2.94 & 8.06  & 2.53   & 13.53 & 11.39      \\
Stage 3                    & 1.82 & 15.34 & 1.52   & 18.68 & 2.32 & 9.70  & 2.42   & 14.45 & 11.22      \\
\bottomrule
\end{tabular}}
\label{t3}
\end{table}

\section{Conclusion}

We propose a three-stage modular speaker diarization system leveraging spatial cues from multi-channel data. 
In the clustering phase, We use the overlap detection model to detect overlapping and non-overlapping segments. 
Non-overlapping segments are processed using beamforming, while overlapping segments are separated with CSS before clustering, which improves the initialization results from CSD across various metrics.
Next, we initialize cACGMM with the results from the first NSD decoding of the multi-channel data to estimate and correct masks, followed by the second NSD decoding. We then use these results for GSS, discarding segments with one or fewer recognized words to refine clustering accuracy. Finally, the third NSD decoding further refines speaker activities, yielding the best recognition results.



\newpage
\bibliographystyle{IEEEtran}
\bibliography{mybib}
\end{document}